\documentclass{ws-ijmpa}
\usepackage[super,compress]{cite}
\usepackage[breaklinks]{hyperref}
\hypersetup{colorlinks,urlcolor=black,citecolor=black,linkcolor=black,filecolor=black}
\usepackage{breakurl}
\usepackage{graphicx}
\usepackage{subfigure}
\usepackage{verbatim}
\usepackage{amsmath}
\usepackage{amssymb}
\usepackage{float}

\newcommand{\simgt}  {\raisebox{-.6ex}{$\stackrel{\textstyle >}{\sim}$}}

\begin{document}
\markboth{P~F~Harrison, R~Krishnan and W~G~Scott}
{Deviations from TBM using a Model with $\Delta(27)$ Symmetry}

%
\catchline{}{}{}{}{}
%

\title{DEVIATIONS FROM TRIBIMAXIMAL NEUTRINO MIXING USING A MODEL WITH $\Delta(27)$ SYMMETRY}

\author{P~F~HARRISON\footnote{p.f.harrison@warwick.ac.uk} \,\,and R~KRISHNAN\footnote{k.rama@warwick.ac.uk}}

\address{University of Warwick,\\Coventry, CV4 7AL, UK}

\author{W~G~SCOTT\footnote{w.g.scott@rl.ac.uk}}

\address{Rutherford Appleton Laboratory,\\Chilton, Didcot, Oxon, OX11 0QX, UK}

\maketitle


\begin{abstract}
We present a model of neutrino mixing based on the flavour group $\Delta(27)$ in order to account for the observation of a non-zero reactor mixing angle ($\theta_{13}$). The model provides a common flavour structure for the charged-lepton and the neutrino sectors, giving their mass matrices a `circulant-plus-diagonal' form. Mass matrices of this form readily lead to mixing patterns with realistic deviations from tribimaximal mixing, including non-zero $\theta_{13}$. With the  parameters constrained by existing measurements, our model predicts an inverted neutrino mass hierarchy. We obtain two distinct sets of solutions in which the atmospheric mixing angle lies in the first and the second octants. The first (second) octant solution predicts the lightest neutrino mass, $m_3 \sim 29~\text{meV}$ ($m_3 \sim 65~\text{meV}$) and the $CP$ phase, $\delta_{CP} \sim -\frac{\pi}{4}$ ($\delta_{CP} \sim \frac{\pi}{2}$), offering the possibility of large observable $CP$ violating effects in future experiments.

\end{abstract}



\section{Introduction}
\label{sec:modelintro}

The phenomenon of neutrino oscillations is characterised by two large mixing angles, the solar angle ($\text{tan}^2\theta_{12}=0.47_{-0.05}^{+0.06}$~\cite{Solar}) and the atmospheric angle ($\sin^2 2\theta_{23}>0.92$~\cite{Atmospheric}), together with the relatively small reactor mixing angle, $\theta_{13}$. The tribimaximal mixing (TBM) scheme~\cite{TBM}, having $\text{tan}^2\theta_{12}=\frac{1}{2}$ and $\sin^2 2\theta_{23}=\nolinebreak 1$, has proved a useful approximation to the data and a substantive stimulus for model building. In TBM,  $\theta_{13}$ is zero and the $CP$ phase ($\delta_{CP}$) is consequently undefined. However, in 2012 the Daya Bay Reactor Neutrino Experiment~\cite{DayaBay, DayaBay2} ($\sin^22\theta_{13}=0.089\pm 0.010\pm0.005$) and RENO Experiment~\cite{RENO} ($\sin^2 2 \theta_{13} = 0.113 \pm 0.013 \pm 0.019 $) showed that $\theta_{13} \simeq 9^o$. T2K~\cite{T2K}, Double Chooz~\cite{DCHOOZ} and MINOS~\cite{MINOS} experiments also measured consistent non-zero values for $\theta_{13}$.

Numerous models based on discrete flavour symmetries~\cite{REVIEW} like $S_3$, $S_4$, $A_4$ etc.~have been proposed to obtain TBM or approximations to it. In any specific model, higher-order corrections to TBM can be evaluated in a systematic manner given the flavour symmetries involved and the field content. In general, all three mixing angles receive corrections of the same order of magnitude~\cite{AF}. Given that the experimentally allowed deviation of $\theta_{12}$ from its TBM value
is much smaller than the measured value of $\theta_{13}$, generating both using higher order corrections is not easy, and has been achieved only in special cases~\cite{Yin, Morisi}.
In this paper we construct instead a model which directly gives a modified TBM consistent with recent observations. 

In many models, Yukawa matrices leading to TBM are constructed in a basis in which either the charged-lepton Yukawa 
matrix or the neutrino Yukawa matrix is diagonal. In contrast, when it was originally proposed~\cite{TBM}, TBM was constructed as the product of a trimaximal matrix (with all the 
eigenstates trimaximally mixed) and a $2\times2$ maximal matrix (with $\nu_1$ and 
$\nu_3$ eigenstates bimaximally mixed). In that scenario, the charged-lepton Yukawa matrix is assumed to be $3 \times 3$ circulant,
\begin{equation}\label{eq:tbmmatricesl}
\left(\begin{matrix}p & q & q^*\\
       q^* & p & q\\
       q & q^* & p
\end{matrix}\right),
\end{equation}
where $p$ is real and $q$ is complex. For a $3 \times 3$ circulant matrix, the diagonalising matrix is trimaximal. On the other hand, the neutrino Yukawa matrix is assumed to have the form,
\begin{equation}\label{eq:tbmmatricesn}
\left(\begin{matrix}x & 0 & y\\
       0 & z & 0\\
       y & 0 & x
\end{matrix}\right),
\end{equation}
where $x$, $y$ and $z$ are real parameters. Such a matrix, which features an embedded $2 \times 2$ circulant, leads to the $2\times2$ maximal diagonalising matrix. 

In the Standard Model, the charged-lepton Yukawa matrix couples the left-handed and the right-handed charged-lepton fields. Similarly, in a minimal extension of the Standard Model, we may construct the Yukawa matrix in the neutrino sector by coupling the left-handed and the right-handed neutrino fields. Here we adopt such a construction which leads to the so called Dirac neutrinos. Moreover, in our model, both the charged-lepton and the neutrino sectors are assumed to have the same flavour structure, i.e.~both the Yukawa matrices have the same form. We propose that the Yukawa matrices ($Y$) comprise the sum of a circulant and a diagonal part (hence `circulant-plus-diagonal'):
\begin{align}
Y&=a I+
\left(\begin{matrix}-\delta -\epsilon & b & b^*\\
       b^* & 2\delta & b\\
       b & b^* & -\delta+\epsilon
\end{matrix}\right).\label{eq:commonyukawa}
\end{align}
We assume $Y$ to be Hermitian, i.e.~$a$, $\delta$, $\epsilon$ are real and $b$ is complex. We remark that pure circulant ($\delta=\epsilon=0$) and pure diagonal ($b=0$) Yukawa matrices allow the construction of the two extremes of the mixing spectrum, namely trimaximal mixing and no mixing respectively. The charged-lepton Yukawa matrix ($Y_l$) and the neutrino Yukawa matrix ($Y_\nu$) of the form, Eq.~(\ref{eq:commonyukawa}), are modelled using the parameters $a_l$, $b_l$ $\delta_l$, $\epsilon_l$ and $a_\nu$, $b_\nu$ $\delta_\nu$, $\epsilon_\nu$ respectively. In the following discussion we consider only the traceless part of the matrix in Eq.~(\ref{eq:commonyukawa}), since the part proportional to the identity ($a I$) does not affect the diagonalising matrix and thus does not affect the mixing.

In the case of the charged leptons, we impose the condition:
\begin{equation}\label{eq:trimaxCondition}
\epsilon_l, \delta_l<< \text{Im}(b_l) << |b_l|. 
\end{equation}
This ensures that $Y_l$ is approximately $3\times 3$ circulant and thus its diagonalising matrix is close to the trimaximal form. For the neutrinos, on the other hand, we impose the condition:
\begin{equation}\label{eq:twomaxCondition}
\text{Im}(b_\nu) << \epsilon_\nu << |b_\nu| << \delta_\nu.
\end{equation}
To analyse the implication of Eq.~(\ref{eq:twomaxCondition}), consider the various $2\times2$ matrices contained in the neutrino Yukawa matrix ($Y_\nu$):
\begin{align}
Y_{\nu (13)}&= \left(\begin{matrix}-\delta_\nu -\epsilon_\nu & b_\nu^*\\
       b_\nu & -\delta_\nu+\epsilon_\nu
\end{matrix}\right), \notag \\
Y_{\nu(12)}&=\left(\begin{matrix}-\delta_\nu -\epsilon_\nu & b_\nu\\
       b_\nu^* & 2\delta_\nu
\end{matrix}\right), \label{eq:yukawaparts} \\
Y_{\nu(23)}&=\left(\begin{matrix} 2\delta_\nu & b_\nu\\
       b_\nu^* & -\delta_\nu+\epsilon_\nu
\end{matrix}\right). \notag
\end{align}
The condition $b_\nu \approx b_\nu^*$ and $\epsilon_\nu << b_\nu$ implies that $Y_{\nu (13)}$ is approximately equal to the $(13)$ submatrix of Eq.~(\ref{eq:tbmmatricesn}). Therefore, the diagonalising matrix for $Y_{\nu (13)}$ is approximately $2\times 2$ maximal. For $Y_{\nu (12)}$ and $Y_{\nu (23)}$, the condition $|b_\nu| << \delta_\nu$ implies that their off-diagonal elements are much smaller than the traceless part of their diagonal elements. The resulting diagonalising matrices of $Y_{\nu (12)}$ and $Y_{\nu (23)}$ will be close to the identity. Therefore, by imposing the condition, Eq.~(\ref{eq:twomaxCondition}), in $Y_\nu$ we obtain an approximate $2\times 2$ maximal diagonalising matrix for the neutrinos as required.

\section{The flavour group - $\Delta(27)$}
Consider the mass term in the Standard Model lagrangian $\psi_R^\dagger Y H \psi_L$ and a flavour transformation $\psi_R \rightarrow C \psi_R , \psi_L \rightarrow C \psi_L$ where $C$ is the regular representation of the cyclic group $C_3$ operating in the generation space, 
and  $C \in \{c, c^2, c^3\}$ with 
\begin{equation} \label{eq:generatorc}
c= \left(\begin{matrix}0& 0& 1\\
       1& 0& 0\\
       0& 1& 0
\end{matrix}\right).
\end{equation}

The mass term will remain invariant under the transformation if $Y$ is a circulant matrix because 
\begin{equation}
C Y_{circ} C^\dagger = Y_{circ}. 
\end{equation}
The regular representation of $C_3$ can be diagonalised using the trimaximal mixing matrix, T
\begin{equation}
T c T^\dagger= d 
\end{equation}
where 
\begin{equation} \label{eq:generatord}
d= \left(\begin{matrix}1& 0& 0\\
       0& \bar{\omega}& 0\\
       0& 0& \omega
\end{matrix}\right).
\end{equation}
The diagonal matrix $d$ comprises the irreducible representations of $C_3$; ($\omega$, $\bar{\omega}$, $1$), where $\omega=-1/2+i\sqrt{3}/2$ and $\bar{\omega}=-1/2-i\sqrt{3}/2$. Trivially, if the Yukawa matrix is diagonal, then
\begin{equation}
d Y_{diag} d^\dagger = Y_{diag}. 
\end{equation}
To build a model with circulant-plus-diagonal Yukawa matrices, Eq.~(\ref{eq:commonyukawa}), it is then natural to use the discrete group having $c$ and $d$ as generators and the group thus obtained is $C_3 \times C_3 \rtimes C_3$. This group, also known as $\Delta(27)$, has been used extensively in model building in earlier studies~\cite{Delta27b,Delta27a,Delta27,King,Delta27c,EMa,Ludl}.

$C_3 \times C_3 \rtimes C_3$ is a 27 element group comprising 11 conjugacy classes whereby we have 11 irreducible representations. Two of those are the defining representation $\boldsymbol{3}$ and its conjugate representation $\boldsymbol{\bar{3}}$. The remaining nine representations are 1-dimensional, comprising the trivial representation and 8 others which transform variously under $C_3 \times C_3 \rtimes C_3$ like the irreducible representations of $C_3$. The character table and the decomposition of tensor products of irreducible representations can be found in Ref.~\citen{Delta27}. 

The tensor product of $\boldsymbol{\bar{3}}$ and $\boldsymbol{3}$ gives all the nine \mbox{1-dimensional} representations. Here we briefly discuss the ones that are relevant to us. Let $X=(X_1, X_2, X_3)^T$ and $Y=(Y_1, Y_2, Y_3)^T$ transform as $\boldsymbol{\bar{3}}$ and $\boldsymbol{3}$ respectively. Then, $\boldsymbol{1}\equiv X^TY = X_1 Y_1 + X_2 Y_2 + X_3 Y_3$ remains invariant and hence it is the trivial representation. The 1-dimensional representations $\boldsymbol{1_c}\equiv X^Tc~Y = X_1 Y_3 + X_2 Y_1 + X_3 Y_2$ and $\boldsymbol{\bar{1}_c}\equiv X^Tc^2~Y = X_1 Y_2 + X_2 Y_3 + X_3 Y_1$ remain invariant under the action of the generator $c$ and they transform as the irreducible representations of $C_3$ under the generator $d$. Analogously we also have $\boldsymbol{1_d}\equiv X^Td~Y = X_1 Y_1 + \bar{\omega} X_2 Y_2 + \omega X_3 Y_3$ and $\boldsymbol{\bar{1}_d}\equiv X^Td^2~Y = X_1 Y_1 + \omega X_2 Y_2 + \bar{\omega} X_3 Y_3$. The complete set of singlets obtained from the tensor product of $\boldsymbol{\bar{3}}$ and $\boldsymbol{3}$ are given in Table~1.
{\renewcommand{\arraystretch}{1.2}
\begin{table}[]
\tbl{The singlets obtained from $\boldsymbol{\bar{3}} \times \boldsymbol{3}$.}
{\begin{tabular}{cc}
\toprule
Representation 	&Expression \\
\colrule
 $\boldsymbol{1}$	&$X^TY$\\
 $\boldsymbol{1_c}$		&$X^Tc~Y$\\
 $\boldsymbol{\bar{1}_c}$	&$X^Tc^2~Y$ \\
 $\boldsymbol{1_d}$		&$X^Td~Y$ \\
 $\boldsymbol{\bar{1}_d}$	&$X^Td^2~Y$ \\
 $\boldsymbol{1_{cd}}$	&$X^Tcd~Y$ \\
 $\boldsymbol{\bar{1}_{cd}}$	&$X^T(cd)^2~Y$ \\
 $\boldsymbol{1_{c\bar{d}}}$	&$X^Tcd^2~Y$ \\
 $\boldsymbol{\bar{1}_{c\bar{d}}}$&$X^T(cd^2)^2~Y$ \\
\botrule
\end{tabular}}
\end{table}}

\section{The Flavour Model}
We construct the present model in a Standard Model framework with Dirac neutrinos, although it can of course be readily extended to a beyond the Standard Model theory having Dirac neutrinos. We assume that the left-handed doublets $(L_e,L_\mu,L_\tau)$ and the right-handed charged leptons $(e_R,\mu_R,\tau_R)$ belong to the representation $\boldsymbol{3}$. Using the $\boldsymbol{\bar{3}}$ $(L_e^\dagger,L_\mu^\dagger,L_\tau^\dagger)$ and the $\boldsymbol{3}$ $(e_R,\mu_R,\tau_R)$, we construct the terms $T_{l}=(L^{\dagger}_e e_R +L^{\dagger}_\mu \mu_R +L^{\dagger}_\tau \tau_R)$, $T_{lc}=(L^{\dagger}_e \tau_R +L^{\dagger}_\mu e_R +L^{\dagger}_\tau \mu_R)$, $\bar{T}_{lc}=(L^{\dagger}_e \mu_R +L^{\dagger}_\mu \tau_R +L^{\dagger}_\tau e_R)$, $T_{ld}=(L^{\dagger}_e e_R +\bar{\omega} L^{\dagger}_\mu \mu_R +\omega L^{\dagger}_\tau \tau_R)$ and $\bar{T}_{ld}=(L^{\dagger}_e e_R +\omega L^{\dagger}_\mu \mu_R +\bar{\omega} L^{\dagger}_\tau \tau_R)$, which transform as $\boldsymbol{1}$, $\boldsymbol{1_c}$, $\boldsymbol{\bar{1}_c}$, $\boldsymbol{1_d}$ and $\boldsymbol{\bar{1}_d}$ respectively. 

We now introduce two flavon fields $\phi_{lc}$ and $\phi_{ld}$ which transform as $\boldsymbol{1_c}$ and $\boldsymbol{1_d}$ respectively and, using the above fermion terms $T_{l}$, $T_{lc}$ and $T_{ld}$, we construct the invariant mass term,
\begin{equation}\label{eq:leptonmassterm}
\begin{split}
&a_l T_{l}H + y_{lc} \left(\frac{\phi^*_{lc}}{\Lambda} T_{lc}H+ \frac{\phi_{lc}}{\Lambda} \bar{T}_{lc}H\right) + y_{ld} \left(\frac{\phi^*_{ld}}{\Lambda} T_{ld}H+\frac{\phi_{ld}}{\Lambda} \bar{T}_{ld}H\right) \\
& \qquad \qquad \qquad \qquad \qquad \qquad \qquad \qquad \qquad \qquad \qquad \qquad \qquad \qquad + H.C.,
\end{split}
\end{equation}
where $H$ is the Standard Model Higgs doublet, $a_l$, $y_{lc}$, $y_{ld}$  are dimensionless real constants and $\Lambda$ is the cut-off scale.
For the neutrinos we assume a similar Dirac mass term with right-handed neutrino fields $(\nu_{eR}$, $\nu_{\mu R}$, $\nu_{\tau R})$ replacing the right-handed charged-lepton fields $(e_R,\mu_R,\tau_R)$ and $i \sigma_2 H^*$ replacing $H$ where $i \sigma_2$ is the $2\times2$ antisymmetric matrix. We also have the neutrino flavons $\phi_{\nu c}$, $\phi_{\nu d}$ and the real constants $a_\nu$, $y_{\nu c}$, $y_{\nu d}$ defined for the neutrinos. 

The flavons acquire vacuum expectation values which spontaneously break the flavour symmetry
and generate the observed fermion Yukawa matrices. The mass term for the charged-leptons, Eq.~(\ref{eq:leptonmassterm}) results in 
\begin{equation}\label{eq:masswithphases}
Y_l=a_l I +
\left(\begin{matrix} 2 {\mathbf y}_{ld} \cos(\theta_{ld})& {\mathbf y}_{lc} e^{i \theta_{lc}}& {\mathbf y}_{lc} e^{-i \theta_{lc}}\\
      {\mathbf y}_{lc}e^{-i \theta_{lc}}&2{\mathbf y}_{ld} \cos(\frac{2\pi}{3}+\theta_{ld})& {\mathbf y}_{lc}e^{i \theta_{lc}}\\
      {\mathbf y}_{lc}e^{i \theta_{lc}}& {\mathbf y}_{lc}e^{-i \theta_{lc}}& 2{\mathbf y}_{ld} \cos(\frac{-2\pi}{3}+\theta_{ld})
\end{matrix}\right)
\end{equation}
where ${\mathbf y}_{lc}=y_{lc} |\langle\phi_{lc}\rangle|/\Lambda$, ${\mathbf y}_{ld}=y_{ld} |\langle\phi_{ld}\rangle|/\Lambda$, $\theta_{lc}=\text{Arg}\langle\phi_{lc}\rangle$, $\theta_{ld}=\text{Arg}\langle\phi_{ld}\rangle$, with $\langle\rangle$ representing a VEV. It can be shown that $Y_l$ is in the circulant-plus-diagonal form, Eq.~(\ref{eq:commonyukawa}). Comparing Eq.~(\ref{eq:commonyukawa}) with Eq.~(\ref{eq:masswithphases}) we obtain the following relationships among the parameters $b_l$, $\delta_l$, $\epsilon_l$ and ${\mathbf y}_{lc}$, $\theta_{lc}$, ${\mathbf y}_{ld}$, $\theta_{ld}$:
\begin{align}
&\begin{aligned}
&\,\,b_l={\mathbf y}_{lc} e^{i \theta_{lc}}
\end{aligned}\\
&\left.
\begin{aligned}
    &\delta_l= {\mathbf y}_{ld}\,\cos (\text{\footnotesize{\(\frac{2\pi}{3}\) }}+\theta_{ld})\\
    &\epsilon_l= -\sqrt{3}\,{\mathbf y}_{ld}\,\sin (\text{\footnotesize{\(\frac{2\pi}{3}\) }}+\theta_{ld})
\end{aligned}
\right\} \quad (\delta_l-\frac{i}{\sqrt{3}} \epsilon_l)\,\bar{\omega}={\mathbf y}_{ld} e^{i \theta_{ld}}.
\end{align}
Similarly, using the parameters $a_\nu$, ${\mathbf y}_{\nu c}$, ${\mathbf y}_{\nu d}$, $\theta_{\nu c}$, $\theta_{\nu d}$, we generate a circulant-plus-diagonal Yukawa matrix ($Y_\nu$) for the neutrinos as well. 

To ensure that the Yukawa matrices have the circulant-plus-diagonal hermitian form, Eq.~(\ref{eq:commonyukawa}), the coupling constants ($a_l$, $y_{lc}$, $y_{ld}$ in the charged-lepton Yukawa term, Eq.~(\ref{eq:leptonmassterm}), and $a_\nu$, $y_{\nu c}$, $y_{\nu d}$ in the corresponding neutrino Yukawa term) need to be real. These constants being real implies an extra symmetry: the Yukawa terms are invariant under the conjugation of the fermion and the flavon fields. Note that we still obtain complex Yukawa matrices because the flavons acquire complex VEVs through spontaneous symmetry breaking. 

\begin{figure}[]
\begin{center}
\includegraphics[scale=0.6]{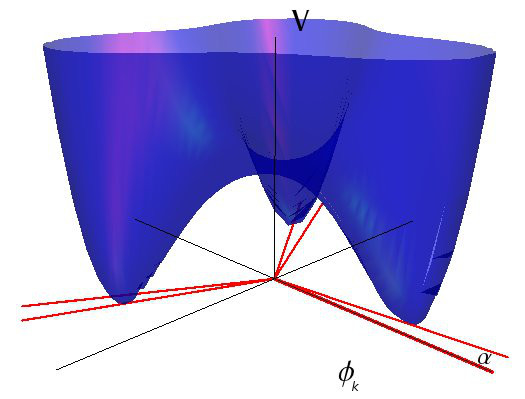}
\caption[]{The flavon potential}
\label{fig:pot}
\end{center}
\end{figure}

For the flavons $\phi_{lc}$, $\phi_{ld}$, $\phi_{\nu c}$ and $\phi_{\nu d}$, we introduce a potential
\begin{equation}\label{eq:potential2}
\text{V}=\sum_{k=lc,~ld,~\nu c,~\nu d} \frac{1}{\Lambda^2}\left|\phi_k^3 - \left(v e^{i\alpha}\right)^3\right|^2
\end{equation}
where $v$ is real, $\alpha<<1$. This potential is invariant under the transformation $\phi_k \rightarrow e^{i \frac{2\pi}{3}n}~\phi_k$, $n$ being an integer. Note that, under the flavour group $\Delta(27)$, the flavons transform as the irreducible representations of $C_3$, i.e.~$\phi_k \rightarrow e^{i \frac{2\pi}{3}n}~\phi_k$. Therefore the potential, Eq.~(\ref{eq:potential2}), is invariant under the flavour group as required. It is also clear that the potential is a positive valued function except when all the terms in the summation go to zero. There are three minima, corresponding to $\phi_k$ equal to $v e^{i\alpha}$ , $v e^{i (\alpha +\frac{2 \pi}{3})}$ and $ve^{i (\alpha-\frac{2 \pi}{3})}$ as shown in Figure \ref{fig:pot}.

The potential is also invariant under the permutation of the index $k$. In other words, the potential ``looks'' the same in relation to all four flavons. This extra permutation symmetry makes the parameter $v e^{i\alpha}$ common to all flavons and the angle $\alpha$ appears in the phases of their VEVs i.e.~$\theta_k=\frac{2\pi}{3}n_k+\alpha$ where $k=lc,~ld,~\nu c,~\nu d$ and $n_k$ are integers. This construction reduces the number of free parameters and renders the model predictive, but even so, it can fit the data, as we see below. The parameter $v$, meanwhile, being common to all the flavons, is irrelevant phenomenologocally because we always have real free parameters ($y_k$) appearing alongside the flavons ($\phi_k$), Eq.~(\ref{eq:leptonmassterm}).

It should be noted that we have not attempted the construction of the most general potential utilising all possible invariant terms. The potential given in Eq.~(\ref{eq:potential2}) merely demonstrates the symmetries involved. In fact, any function of the flavons which satisfies $\Delta(27)$ flavour symmetry as well as the earlier mentioned permutation symmetry and which is bounded from below, is a suitable candidate for the potential. Such a function will have a set of minima which correspond to all flavons having three complex values separated by a phase of $\frac{2\pi}{3}$ i.e.~of the form $v e^{i\alpha}$ , $v e^{i (\alpha +\frac{2 \pi}{3})}$ and $ve^{i (\alpha-\frac{2 \pi}{3})}$. Therefore the specific structure of the flavon potential is irrelevant to the present discussion.

A small but non-zero $\alpha$, along with the integer $n_{lc}=0$ is a necessary condition to obtain the relation $\text{Im}(b_l)<<|b_l|$  as given in Eq.~(\ref{eq:trimaxCondition}). Similarly, for the case of the neutrinos, the smallness of $\alpha$ ensures $\text{Im}(b_\nu)<<|b_\nu|$, as given in Eq.~(\ref{eq:twomaxCondition}). A small $\alpha$ also leads to the condition $\epsilon << \delta$, Eq.~(\ref{eq:twomaxCondition}), since $\alpha \rightarrow 0$ makes two diagonal elements equal. Clearly the extreme limit 
\begin{equation}\label{eq:masterInequal}
\frac{{\mathbf y}_{ld}}{{\mathbf y}_{lc}}<<\alpha<<\frac{{\mathbf y}_{\nu c}}{{\mathbf y}_{\nu d}} << 1 \end{equation}
leads to TBM\footnote{We do not provide a mechanism to explain the hierarchical constraints, Eq.~(\ref{eq:masterInequal}), among the free parameters of the model. This could be taken up as a topic for future research.}. The fact that Eq.~(\ref{eq:masterInequal}) results in a mixing close to TBM can also be varified numerically. The deviations from this limiting TBM mixing are calculable, in principle, in terms of the ``small'' numbers, $\frac{{\mathbf y}_{ld}}{{\mathbf y}_{lc}}$, $\alpha$ and $\frac{{\mathbf y}_{\nu c}}{{\mathbf y}_{\nu d}}$. Overall, the model has 7 independent parameters, $a_l$, ${\mathbf y}_{lc}$, ${\mathbf y}_{ld}$, $a_\nu$, ${\mathbf y}_{\nu c}$, ${\mathbf y}_{\nu d}$ and $\alpha$, which determine 10 observables. This constrains the parameter space of the masses and the mixing observables. In the next section we fit these parameters to the 7 experimentally measured observables (three charged-lepton masses, two neutrino mass-squared differences and the solar and the reactor mixing angles). We obtain solutions for the atmospheric mixing angle in both the first and the second octants and make predictions for the observables yet to be measured (the lightest neutrino mass and the $CP$ phase).

\section{Fitting the model to experimental data}

We use the masses of the charged leptons and the mass-squared differences of the neutrinos, renormalised at 1~TeV, from Ref.~\citen{XingMass}: 
\begin{align}\label{eq:exptMasses}
m_e&=0.4959~\text{MeV};~m_\mu=104.7~\text{MeV};~m_\tau=1780~\text{MeV};\\ 
&m_2^2 - m_1^2=91~\text{meV}^2;~|m_3^2 - m_2^2|=2900~\text{meV}^2.
\end{align}
Since the mass ratios change only very slowly under renormalisation evolution~\cite{Invariants}, the fit will remain valid not only at 1~TeV, but also at the unknown scale $\Lambda$, provided that lies within a few orders of magnitude of 1~TeV. After fixing the charged-lepton masses and the neutrino mass-squared differences as above (5 observables), the charged-lepton and neutrino mass matrices (7 parameters) were generated by Monte Carlo (using 7-5=2 random variables). From the generated mass matrices we computed each time the PMNS matrix. Experimental values for the solar and the reactor mixing angles were taken from Ref.~\citen{NeutrinoGlobalFit}
\begin{equation}\label{eq:exptMixings}
\sin^2\theta_{12}= 0.313_{-0.012}^{+0.013};~\sin^2\theta_{13}=0.0252_{-0.0023}^{+0.0022}.
\end{equation}
These data are compared with the values extracted from the Monte Carlo generated PMNS matrices using a $\chi^2$ goodness of fit variable:
\begin{equation}\label{eq:chi2}
\chi^2 = \displaystyle\sum_{\theta=\theta_{12},\theta_{13}}\left(\frac{(\sin^2\theta)_{\text{model}}-(\sin^2\theta)_{\text{expt}}}{\sigma_{\text{expt}}}\right)^2
\end{equation}
where $\sigma_{\text{expt}}$ is the experimental error on $\sin^2\theta$.

\begin{figure}[]
\begin{center}
\includegraphics[scale=1.5]{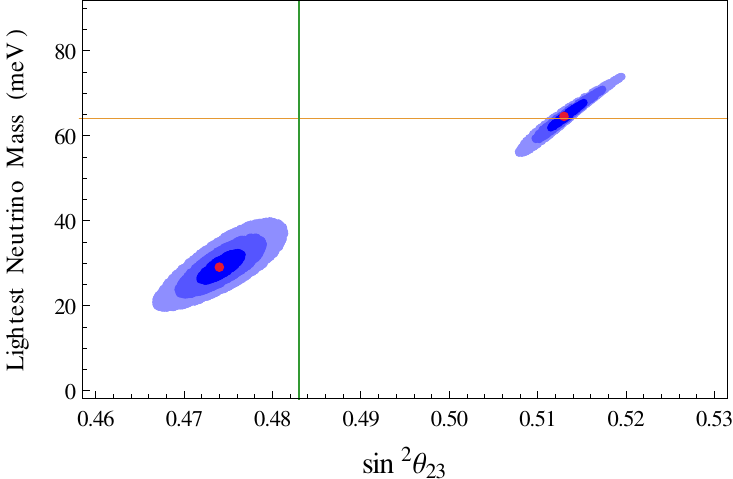}
\caption[Predictions made by the model]{Predicted values of the mass of the lightest neutrino mass eigenstate~vs.~$\sin^2 \theta_{23}$ showing solutions in both first and second octants. The three shades of blue denote $\chi^2\leq9$, $\chi^2\leq4$, $\chi^2\leq1$ and the best fits are indicated by the red dots. The orange line corresponds to the mass limit, $\sum_i m_{\nu_i} < 0.23~\text{eV}$, imposed by the Planck data. The green line is the upper limit of $\sin^2 \theta_{23} = 0.436_{-0.032}^{+0.047}$ ($1\sigma$ errors) from the global fit~\cite{NeutrinoGlobalFit}.}
\label{fig:neutrinopredict1}
\end{center}
\end{figure}

\begin{figure}[]
\begin{center}
\includegraphics[scale=1.5]{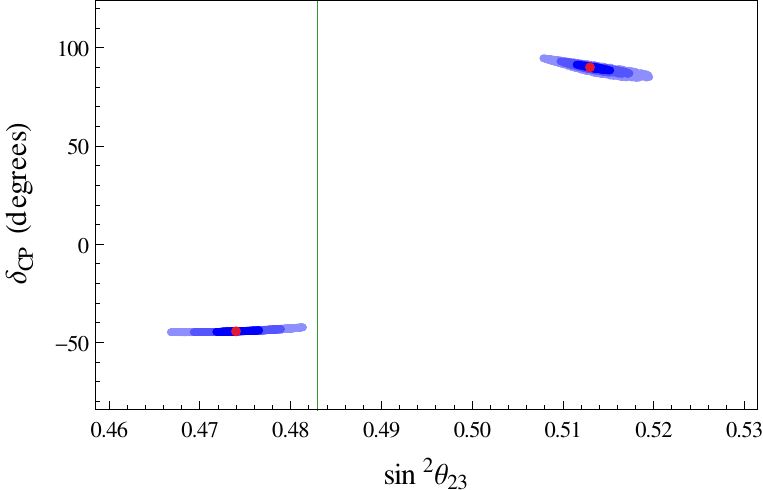}
\caption[Predictions made by the model]{Predicted values of $\delta_{CP}$~vs.~$\sin^2 \theta_{23}$. The first and the second octant solutions correspond to $\delta_{CP} \sim -\frac{\pi}{4}$ and $\delta_{CP} \sim \frac{\pi}{2}$ respectively. The green line is the upper limit of $\sin^2 \theta_{23} = 0.436_{-0.032}^{+0.047}$ ($1\sigma$ errors) from the global fit~\cite{NeutrinoGlobalFit}.}
\label{fig:neutrinopredict2}
\end{center}
\end{figure}

The results of the $\chi^2$ analysis are shown in Fig.~\ref{fig:neutrinopredict1} and Fig.~\ref{fig:neutrinopredict2}. With its parameters thus determined by the values of the observables given in Eqs.~(\ref{eq:exptMasses}) and (\ref{eq:exptMixings}), our model predicts two sets of solutions corresponding to the octant degeneracy of the atmospheric mixing angle. A part of the second octant solution is excluded by the recent measurement of the sum of the neutrino masses, $\sum_i m_{\nu_i} < 0.23~\text{eV}$, by the Planck satellite~\cite{Planck}. Among the 10 observables (6 masses and 4 mixing observables), the lightest neutrino mass (or, equivalently, the overall neutrino mass offset) and the $CP$-violating phase, $\delta_{CP}$, are largely unknown. So the model is used to predict these quantities. We note that our numerical analysis does not give an acceptable $\chi^2$ for any normal hierarchy, whereby an inverted hierarchy can be said to be a prediction of our model, given the measured observables.

In the first octant, the best fit\footnote{The best fit points are given by  $a_l=628.3$, ${\mathbf y}_{lc}=575.7$, ${\mathbf y}_{ld}=31.7$, $a_\nu=9.4$, ${\mathbf y}_{\nu c}=13.9$, ${\mathbf y}_{\nu d}=-33.7$, $\alpha=-2.40^\circ$ for the first octant and $a_l=628.3$, ${\mathbf y}_{lc}=576.5$, ${\mathbf y}_{ld}=9.2$, $a_\nu=21.3$, ${\mathbf y}_{\nu c}=7.9$, ${\mathbf y}_{\nu d}=-52.4$, $\alpha=-2.94^\circ$ for the second octant, along with the integers $n_{l c}, n_{l d}, n_{\nu c}, n_{\nu d}$ equal to $0, 1, 1, -1$ respectively. The values of $a_l$, ${\mathbf y}_{lc}$, ${\mathbf y}_{ld}$ are in units of $\text{MeV}/h_o$ and $a_\nu$, ${\mathbf y}_{\nu c}$, ${\mathbf y}_{\nu d}$ are in $\text{meV}/h_o$ where $h_o$ is the Standard Model Higgs VEV. Our fit corresponds to the charged-lepton masses $m_e\simeq 0.5~\text{MeV}$, $m_\mu\simeq 105~\text{MeV}$ and $m_\tau\simeq 1780~\text{MeV}$ as required.} gives the neutrino masses, $m_3\simeq 29.0~\text{meV}$, $m_1\simeq 60.4~\text{meV}$ and $m_2\simeq 61.2~\text{meV}$. The moduli-squared of the elements of the PMNS matrix corresponding to the best fit are:
\begin{equation}
|\text{U}_\text{PMNS}|^2 = \left(\begin{matrix}0.669 & 0.305 & 0.025\\
       0.226 & 0.312 & 0.462\\
       0.105 & 0.382 & 0.513
\end{matrix}\right).
\end{equation}
This gives $\sin^2\theta_{12}=0.313$, $\sin^2\theta_{23}=0.474$, $\sin^2\theta_{13}=0.025$ and $\delta_{CP}=-44.5^\circ$ with the Jarlskog $CP$-violating invariant, $J=-0.025$ (to be compared with its theoretical maximum value, $J_{\text{max}}=\frac{1}{6\sqrt{3}}\simeq 0.096$).

The best fit\footnotemark[\value{footnote}] in the second octant gives $m_3\simeq 64.5~\text{meV}$, $m_1\simeq 83.5~\text{meV}$, $m_2\simeq 84.1~\text{meV}$, $\sin^2\theta_{12}=0.312$, $\sin^2\theta_{23}=0.513$, $\sin^2\theta_{13}=0.025$, $\delta_{CP}=90.0^\circ$ and $J=0.036$.

\section{Conclusion}

In general, models that produce TBM can explain only a small non-zero reactor angle ($\theta_{13}$) by introducing higher order corrections. In this paper we show that Yukawa matrices containing circulant and diagonal parts are well-adapted for constructing neutrino mixing patterns with potentially large deviations from TBM. The discrete group $\Delta(27)$ constructed using circulant (Eq.~\ref{eq:generatorc}) and diagonal (Eq.~\ref{eq:generatord}) generators is a natural choice to construct such circulant-plus-diagonal Yukawa matrices. The potential term of the flavons has been assumed to have an extra permutation symmetry and this leads to a 7-parameter model. We have used this model to fit present neutrino oscillation data, including the non-zero reactor angle. Once constrained by these measurements, our model predicts a neutrino mass spectrum of the inverted hierarchy ($m_3^2 < m_1^2 < m_2^2$). For $\theta_{23}$ in the first octant, we predict the lightest neutrino mass, $m_3 \sim 29~\text{meV}$, and a $CP$ phase, $\delta_{CP}\sim-\frac{\pi}{4}$ and for $\theta_{23}$ in the second octant we predict $m_3 \sim 64~\text{meV}$ and $\delta_{CP}\sim\frac{\pi}{2}$. It should be noted that, unlike the vast majority of flavour models in the literature, here we use only singlet flavons to obtain the desired mixing pattern.

We have checked that the given $CP$ phases are not a generic prediction arising from the structure of the model itself. Rather, the value of $\delta_{CP}$ depends on several of the parameters and these are constrained here by the measured values of other observables. In fact $\delta_{CP}$ is strongly correlated with $\theta_{13}$ and comes out close to the limiting values of $\frac{\pi}{2}$ and $-\frac{\pi}{4}$ when $\theta_{13}$ becomes relatively large, $\theta_{13}\, \simgt\,\, 0.1$. This offers the potential for large observable $CP$-violating asymmetries in future neutrino oscillation experiments, with Jarlskog's $CP$-violating invariant, $J$, assuming close to 40\% and 25\% of its theoretical maximum value.

\section*{Acknowledgments}

PFH acknowledges support from the UK Science and Technology Facilities Council (STFC) on the STFC consolidated grant ST/H00369X/1. Two of us (PFH and RK) acknowledge the hospitality of the Centre for Fundamental Physics (CfFP) at the Rutherford Appleton Laboratory. We also acknowledge useful comments by Steve King.

\end{document}